\documentclass[12pt,letterpaper]{article}          
\usepackage{osajnl}
\usepackage[draft]{hyperref} 
\usepackage{epstopdf}

\begin{document}

\title{
II. A method of estimating time scales of atmospheric piston and its application at Dome\,C (Antarctica).}

\author{A. Kellerer, M. Sarazin}\address{European Southern Observatory}\email{aglae.kellerer@eso.org}
\author{T. Butterley, R. Wilson}\address{Department of Physics, University of Durham}

\begin{abstract}
Temporal fluctuations of the atmospheric piston are critical for 
interferometers as they determine their sensitivity.
We characterize an instrumental set-up, termed the piston scope, 
that aims at measuring the atmospheric time constant, $\tau_0$,
through the image motion in the focal plane of a Fizeau interferometer.\\
High-resolution piston scope measurements have been obtained at the observatory of Paranal, Chile, in April 2006. 
The derived atmospheric parameters are shown to be consistent with 
data from the astronomical site monitor, 
provided that the atmospheric turbulence is displaced along a single direction.\\
Piston scope measurements, of lower temporal and spatial resolution, 
were for the first time recorded in February 2005
at the Antarctic site of Dome\,C.
Their re-analysis in terms of the new data calibration sharpens the conclusions of
a first qualitative examination [\ref{Kellerer1}].

\end{abstract}

\ocis{120.0120, 010.0010}

\maketitle 

\section{Introduction}
Interferometers have been introduced in astronomy to gain in spatial resolution without 
the need to build extremely large telescopes.
To resolve {\it Sirius\/}, observations in the infrared domain ($\sim2\,\mu$m) would
require a telescope of about 170\,m mirror diameter.
Fortunately, {\it Sirius \/} can also be resolved by two telescopes of more modest size, 
separated by 170\,m and operated as an interferometer.
Yet, despite this considerable gain in resolution, interferometers are not the prime tool of today's astronomers.
This is largely due to their limited sensitivity: atmospheric turbulence makes the interferometric fringe pattern move in the detector plane.
Accordingly, one tends to use exposure times that are short enough to ``freeze" the turbulence, i.e. typically several milliseconds.
To increase the sensitivity, phasing devices are being designed that
measure the position of the fringe pattern due to a reference star, and correct continuously 
for the fringe motion of the target object. 
For such devices to work, a sufficient number of photons need to be collected on the reference star 
during the time when the atmosphere is frozen, i.e. during the {\it atmospheric coherence time\/} $\tau_0=0.314 \; r_0 / \overline{V}_{5/3}$, where $r_0$ is the {\it Fried parameter\/} and $\overline{V}_{5/3}$ is a weighted average of the turbulent layers' velocities.
Clearly, the coherence time is the parameter that determines the performance of today's interferometers. 
Different definitions of the atmospheric coherence time have been introduced  
in relation to various observational techniques:
single telescopes with or without adaptive-optics, 
interferometers with or without fringe trackers etc.
However the standard adaptive-optics coherence time $\tau_0$ has been shown 
to quantify the performance of all these techniques\,[\ref{Kellerer2}].\\
In a previous article [\ref{Kellerer1}], we characterized the temporal evolution of fringe motion at Dome\,C, a summit on the antarctic continent, and a potential site for a future interferometer,
using the motion of the fringe pattern formed in the focal plane of a Fizeau interferometer. 
The temporal and spatial sampling of the measurements were low due to the available equipment and, instead of determining coherence-time values,
the mean duration of correlation  was assessed
by fitting the fringe correlation-function onto an exponential curve (cf. Section\,\ref{DomeC}). 
Such measurements have now been repeated at the site of Paranal, Chile, with sufficient spatial and temporal sampling, to allow the determination of the coherence time. Further, all relevant atmospheric parameters are constantly monitored at Paranal by a meteorological station, hence the parameter values derived through our set-up (termed {\it piston scope\/}) can be checked against reference values.\\
In the first Section, the quantities measured with the piston scope are related to the following atmospheric parameters: the Fried parameter, the turbulent layers velocities and the coherence time -- using the Kolmogorov theory of atmospheric turbulence.
The relations are then tested on the observations performed at Paranal. It is shown that when the sampling is sufficient, the precision on the coherence time is limited by the piston scope's sensitivity to wind direction.
Given these results, the third Section presents a new analysis of the measurements obtained at Dome\,C [\ref{Kellerer1}]. The lower limits to the coherence time, derived through our first qualitative analysis, are confirmed and additional results on the Fried parameter and wavefront speed are given.

\section{Formalism}\label{Formalism}
\begin{figure*}
\centering
\includegraphics[width=10cm]{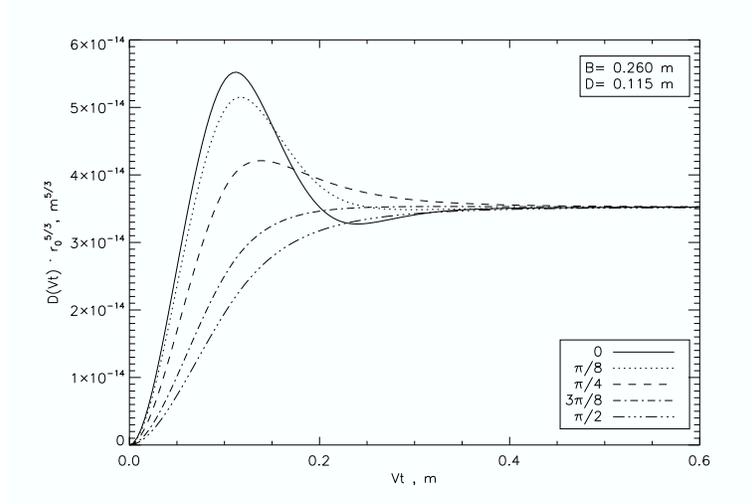}
\caption{Structure functions of the fringe position relative to the combined Airy discs, for an interferometer with mirror diameters $D$
and baseline length $B$. The atmosphere is assumed to consist of a single layer displaced with wind speed $V$ at an angle $\alpha$ from the baseline. The values of $\alpha$ are indicated in the bottom right box.
}
\label{Fig:SF}
\end{figure*}

The purpose of the piston scope experiment is to track the rapid fluctuations of the atmospheric piston. 
To this effect, the entrance pupil of a telescope
is covered by a mask with two circular openings. The resulting image is a fringe pattern within the superposition of the two diffraction discs.
Atmospheric turbulence keeps the image of the star moving on the detector. The local inclination of the wave front over each of the holes causes the movement of the Airy discs, whereas difference in the optical path for the two holes, i.e. the piston, shifts the fringe pattern relative to the center of the Airy discs. Telescope vibrations, on the other hand, cause merely a common movement of  Airy discs and fringes. The relative movements between Airy discs and fringes are, therefore, solely due to the atmospheric turbulence. The subsequent analysis deals with their temporal patterns.
Piston changes shift the fringe pattern relative to the Airy discs along the interferometric axis. Accordingly,  in order to assess the temporal fluctuations of the piston it is sufficient to consider the shift along the axis. \\
As suggested by Conan et al. [\ref{Conan}], the spatial power spectrum $W_\phi$ of the relative movements between Airy discs and fringes is  derived from the phase spectrum $W_{\varphi}$, 
assuming a Kolmogorov model of turbulence with an infinite outer scale. 
In the following we use the notations of Conan et al. [\ref{Conan}].
\begin{eqnarray}
W_{\varphi}({\bf f}) = 
0.00969 \; k^2
\int_0^{+\infty}  f^{-11/3} \;  C_{\rm n\/}^2 \; {\rm d\/}h ,
\label{eq:Wphi}
\end{eqnarray}
where ${\bf  f}$ is the  spatial frequency and 
$  k = 2  \pi /\lambda$ the wavenumber.
The turbulence intensity of a layer $i$ of thickness d$h$
at altitude $h$ is specified in terms of $C_{\rm n\/}^2\; {\rm
d}h$.  
The explicit  dependence of $C_{\rm n\/}$ and all following parameters on $h$ is dropped
to ease  the  reading of  the formulae.  The measured quantity is the separation -- along the interferometric axis, $x$ -- between the central fringe and the center of the combined Airy discs. 
The spatial filter $\tilde{M}$ that converts  $W_{\varphi}$ into  the
power spectrum $W_\phi$ equals:
\begin{eqnarray}
\label{eq:Mphi}
\tilde{M} ({\bf f}) &=& \lambda / (2\pi) \; A({\bf f})\; FT[(\delta_{\rm B\/}-\delta_0)/B - (\delta_{\rm B\/}+\delta_0)/2\ast {\rm d\;/d\/}x] ({\bf f}),
\end{eqnarray}
for  a baseline  vector ${\bf  B}$  and the  aperture filter  function
$A({\bf f})$.  For a circular aperture of diameter $D$, $A ({\bf f}) =
2 J_{1}(\pi fD)/  (\pi fD)$ and $f = | {\bf  f}|$. $J_{\rm n\/}$ stands
for the Bessel function of order $n$. 
$FT$ represents the Fourier transform, $\delta_{\rm L\/}$ is the delta function centered on $L$ and
$\ast$ denotes convolution. Hence,
\begin{eqnarray}
\tilde{M} ({\bf f}) &=&  \lambda / (2\pi) \; A({\bf f})\;[2 \; \sin ( \pi {\bf fB}) - 2\pi {\bf fB}\; \cos (\pi {\bf fB}) ]\;/\;B  \\
W_\phi ({\bf f}) &=& \tilde{M}^2({\bf f}) \; W_{\varphi}({\bf f}).
\end{eqnarray}
In the single layer approximation,  we assume the turbulent layer to be transported  with a
velocity ${\bf  V\/}$ directed at an  angle $\alpha$ with  respect to the
baseline. The temporal  power spectrum of the measured quantity  is obtained
by integrating in the frequency plane  over a line displaced by $f_{\rm x\/} =
\nu/V$ from the coordinate origin  and inclined at angle $\alpha$. Let
$f_{\rm y\/}$  be  the  integration  variable  along  this  line  and  $f^2  =
f_{\rm x\/}^2+f_{\rm y\/}^2$.  The temporal power spectrum  equals:
\begin{eqnarray}
\label{eq:w-nu}
w_\phi (\nu) &=& \frac{1}{V} \int_{- \infty}^{+\infty} 
W_\phi \left( f_{\rm x\/} \cos \alpha +  f_{\rm y\/} \sin \alpha , f_{\rm y\/} \cos \alpha  - f_{\rm x\/} \sin \alpha
\right) \; {\rm d\/}f_{\rm y\/}  
\end{eqnarray}
We then derive the expression of the structure function:
\begin{eqnarray}
D_\phi (t) &=& 2 \int_{- \infty}^{+\infty}  (1-\cos(2 \pi \nu t)) w_\phi (\nu) {\rm d\/}\nu \\
&=& 2 \times 0.00969 \; C_{\rm n\/}^2 \; {\rm d\/}h \; / \; B^2 \int_{0}^{+\infty} f^{-8/3} (2 J_{1}(\pi fd)/  (\pi fd))^2 {\rm d\/}f \nonumber \\
&& \int_{0}^{2\pi} (1-\cos(2 \pi f \cos (\theta + \alpha) V t)) \nonumber \\
&& [2 \sin (\pi B f \cos\theta)-2\pi f B \cos \theta \cos(\pi f B \cos \theta)]^2 
{\rm d\/}\theta 
\label{eq:SF}
\end{eqnarray}
The best estimate of the parameters is obtained by fitting the measured points to: $D_\phi (t) + K$, where $K$ is a constant that allows for white measurement noise.
As seen from Eq.\,\ref{eq:SF}, the structure function depends on the wind orientation $\alpha$
because the mask of the piston scope is  not rotationally symmetric.  
Temporal evolutions of the structure functions, for different values of $\alpha$, are represented on Fig.\,\ref{Fig:SF}.
The asymptotic value of the structure function at large time increments is determined by the Fried parameter $r_0$,
whereas the time needed to reach the asymptotic value is a function of the velocity $V$. 
\section{Measurements at Paranal}\label{Paranal}
\subsection{Observational set-up}
\begin{figure*}
\centerline{
\includegraphics[height=7cm,width=10cm]{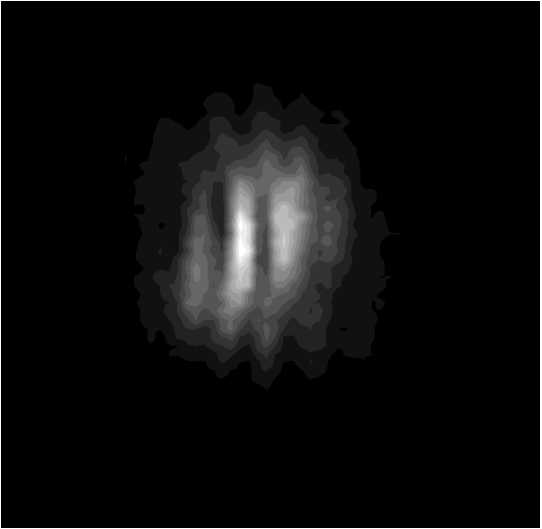}
\includegraphics[height=7cm,width=10cm]{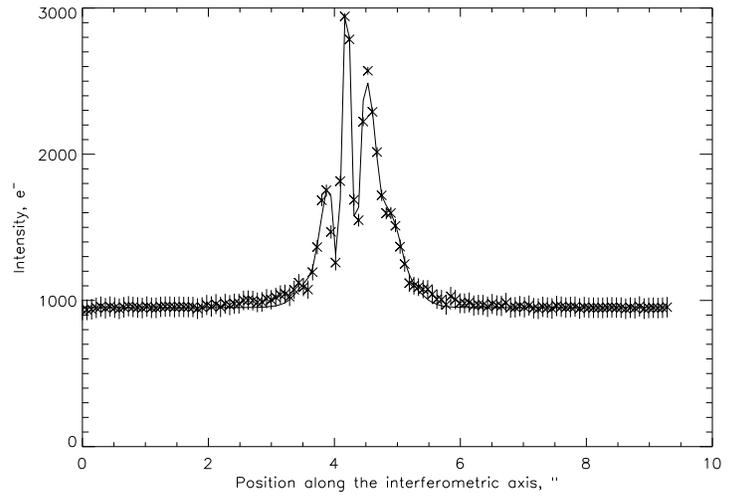}}
\caption{ Example of an image recorded with 1\,ms exposure time 
at Paranal on the night of 23-24 April at 02:03:55 UT
and fitted intensity profile along the axial direction.
}
\label{Fig:Fit}
\end{figure*}

Several observations of {\it Spica\/} were obtained at Paranal on the nights from 22-23 and 23-24 April 2006, using a modified SLODAR [\ref{SLODAR}] (Slope detection and ranging).
This SLODAR is designed  to measure profiles of the atmospheric turbulence
with a telescope that has a 0.4\,m diameter primary mirror, and a focal length of 4.064\,m.
The detector is a $128 \times 128$ array of $(24 \times 24)\mu$m$^2$ pixels with a peak quantum efficiency of 92\% at $\lambda_0=550$\,nm and next to zero read-out noise. 
For our experiment the entrance pupil of SLODAR was covered by a mask with two circular openings of diameter $D=0.115$\,m and centers $B=0.260$\,m apart. The resulting image is a fringe pattern of angular period
$\lambda_0/B=0.44"$ within the superposition of two Airy discs of diameter $2.44\lambda_0/D=2.41"$.
Two lenses were used to increase the focal length by a factor 16.67, this makes each pixel correspond to an angular increment of 0.073".
During the first night, a sequence of 1000 images was recorded at 240\,Hz 
with an exposure time equal to 2\,ms.
On the following night, six sequences of 1000 images were recorded at 
300\,Hz with 1\,ms exposure time. \\
The piston is quantified in terms of the motion of the fringe packet relative to the combined Airy discs.
The quantification of the axial motion requires the extraction of the following parameters from the observed images: the position of the central fringe and the position -- along the interferometric axis -- of the center of the combined Airy discs.
This extraction has been described in detail in a previous article [\ref{Kellerer1}].
An example of a raw image is shown on Fig.\,\ref{Fig:Fit} 
with the corresponding, fitted intensity profile.

\subsection{Derivation of atmospheric parameters}
\begin{figure}
\centering
\includegraphics[width=8cm]{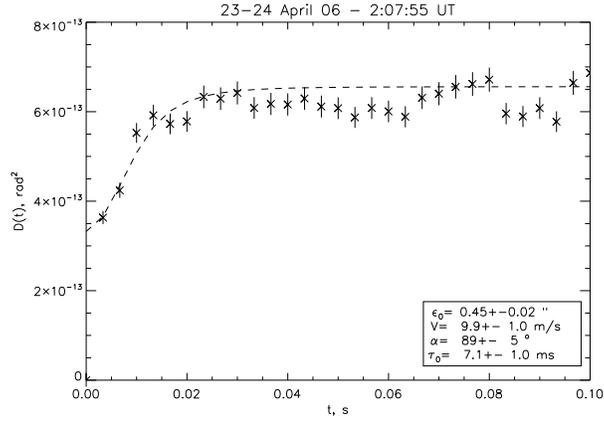}
\caption{
Theoretical structure functions (dashed line) fitted onto data obtained at Paranal,
the resulting seeing $\epsilon_0$, velocity $V$, wind orientation $\alpha$ 
and coherence time $\tau_0$ are indicated.
}
\label{Fig:P}
\end{figure}

The Fried parameter $r_0$, the wavefront velocity $V$ and orientation $\alpha$ are derived 
by fitting $D_\phi (t) + K$ onto the data points, as described in Section\,\ref{Formalism}.
$D_\phi (t)$ corresponds to an atmospheric model where the turbulence is contained in a single layer, 
that is displaced as a whole with the velocity $V$ under an angle $\alpha$.\\
The resulting parameter values and uncertainties are indicated on Fig.\,\ref{Fig:P}. 
The latter correspond to a doubling of the squared deviation of the data points to the theoretic structure function.
The Fried parameter is determined by the asymptotic value of the structure function at large time increments.
To ease the comparison with the meteorological station of Paranal, we indicate the {\it seeing angle\/} 
$\epsilon_0$ rather than the Fried parameter $r_0$, these two parameters are essentially equivalent: $\epsilon_0=0.976\;\lambda / r_0$\,[rad].
$V$ and $\alpha$ are derived from the first few measurement points and
the coherence time, $\tau_0$, is then obtained through the classic relation: $\tau_0  = 0.314\; r_0  / V$.
\subsection{Performance of the piston scope}\label{Sec:Performance}
\begin{figure}[htbp]
\centerline{\includegraphics[width=18cm]{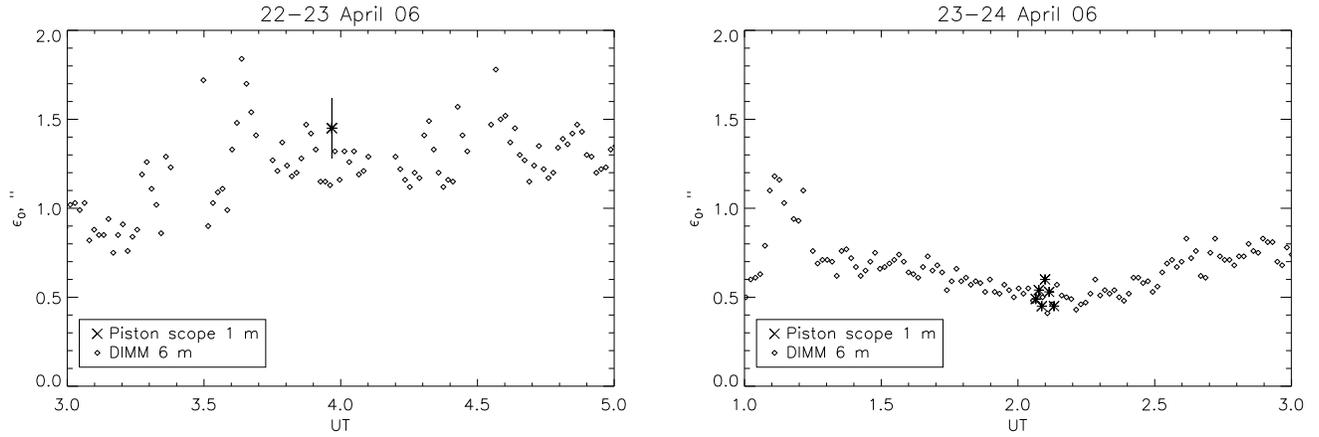}}
\caption{ Seeing values measured at Paranal with the DIMM and the piston scope. 
The uncertainties of the piston scope values correspond to a twofold increase in the quality of the data adjustment.
}
\label{Fig:ASM_s}
\end{figure}

\begin{figure}[htbp]
\centerline{\includegraphics[width=18cm]{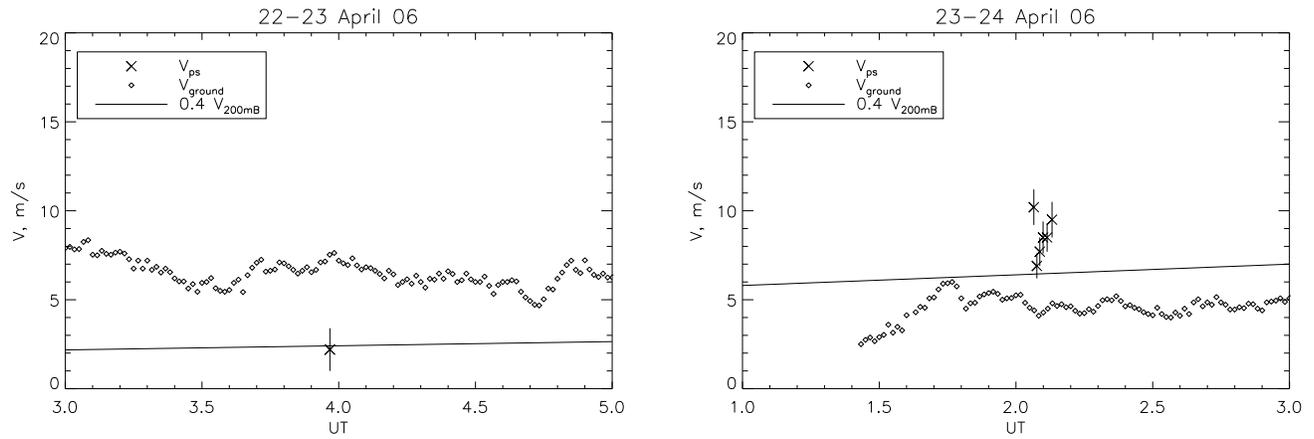}}
\caption{ Wavefront velocities obtained with the piston scope ($V_{\rm ps \/}$), 
wind velocities measured by sensors at 30\,m above the ground of Paranal ($V_{\rm g \/}$)
and interpolated  at 200\,mB from ECMWF data ($V_{\rm 200 mB}$).
}
\label{Fig:ASM_w}
\end{figure}

\begin{figure}[htbp]
\centerline{\includegraphics[width=18cm]{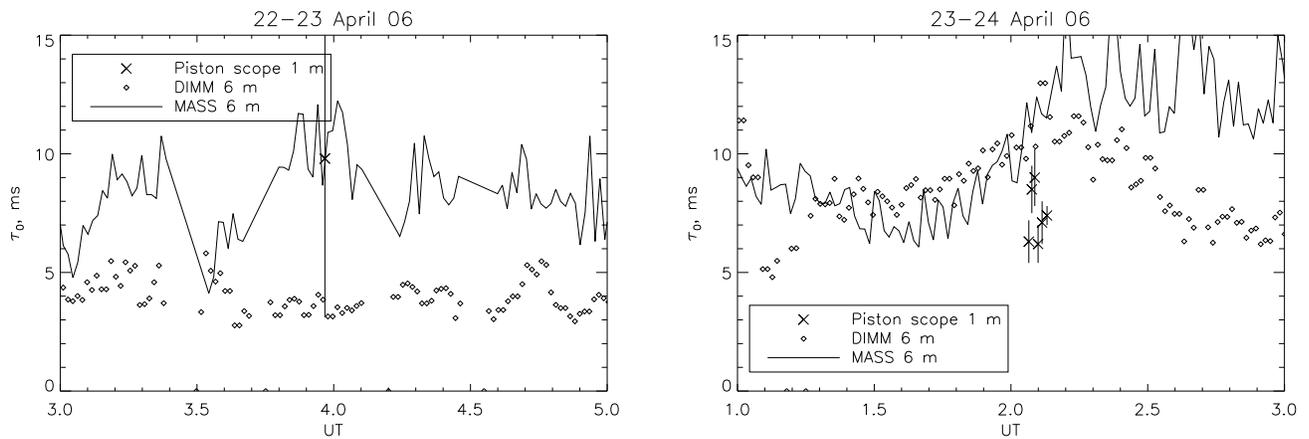}}
\caption{ Coherence times obtained at Paranal through three different methods.
}
\label{Fig:ASM_t}
\end{figure}

\begin{figure}[htbp]
\centerline{\includegraphics[width=18cm]{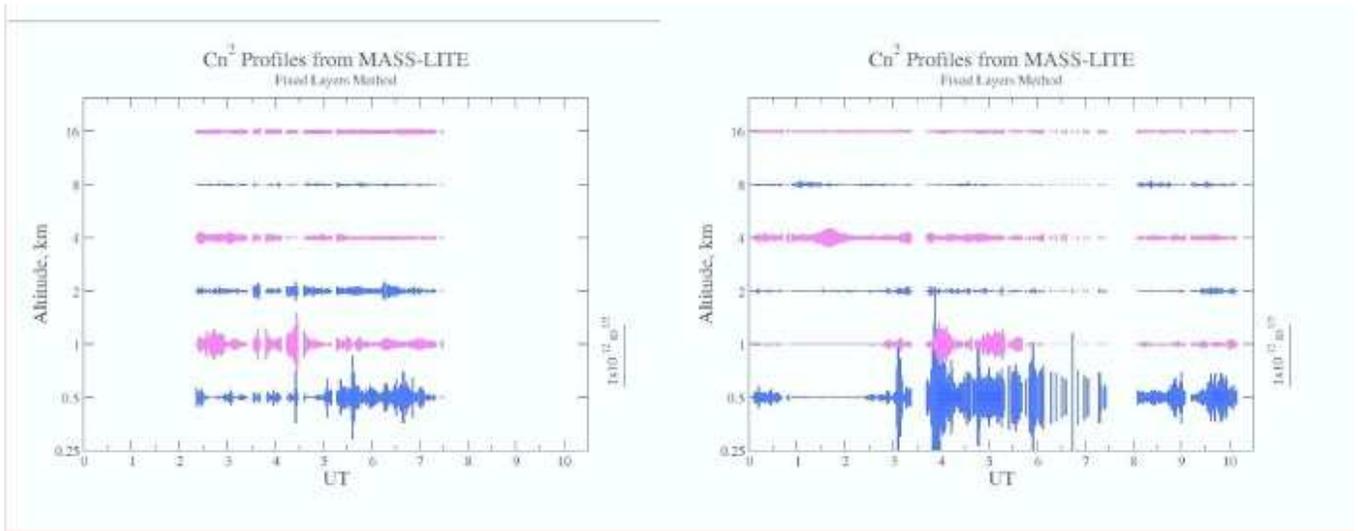}}
\caption{ Profiles of the free atmosphere turbulence obtained by MASS at Paranal.
On 22-23 April (left panel) the turbulence was contained in several layers of similar intensity,
while on 23-34 April (right panel) one layer at 4\,km was predominant around 2:00\,UT.
}
\label{Fig:Profiles}
\end{figure}

On Figs.\,\ref{Fig:ASM_s}-\ref{Fig:ASM_t}, the values of $\epsilon_0, V_{\rm ps \/}$ and $\tau_0$ obtained with the piston scope are compared to measurements in terms of the Paranal monitoring-instruments. 
We do not compare the wind orientations, because the value of $\alpha$ that is obtained with the piston scope 
depends on the position of the mask, hence on the pointing of the telescope, and  
it is difficult to relate it to the angle measured by the meteorological station.\\
\begin{itemize}
\item Seeing values (see Fig.\,\ref{Fig:ASM_s}): 
The values estimated with the piston scope coincide
with those measured at 6\,m height by the 
DIMM [\ref{DIMM}] (Differential Image Motion Monitor).
Note that we assume the atmosphere to consist of one layer
displaced along a single direction, yet Fig.\,\ref{Fig:Profiles} shows that on 22-23 April
the turbulence was contained in several layers with similar intensity.
However, the asymptotic value of the structure function has the same altitude dependence as the seeing:
\begin{equation}
D_\phi (t \gg \tau_0) \propto r_0^{-5/3} \propto \int_{0}^{+\infty} C_{\rm n\/}^2\, {\rm d\/}h
\end{equation}
therefore the seeing estimated by the piston scope is correct independently of 
turbulence profile.

\item Velocities (see Fig.\,\ref{Fig:ASM_w}):
The wavefront velocity $V_{\rm ps \/}$ derived with the piston scope is a turbulence-weighted 
average of the layers' velocities $V(h)$. 
Ideally, $V_{\rm ps \/}$ should have the same dependence on turbulence parameters
as $\tau_0$, hence:
\begin{equation}
V_{\rm ps \/} \propto \overline{V}_{5/3} = [ \frac{\int_{0}^{+\infty} V(h)^{5/3}\;C_{\rm n\/}^2(h)\, {\rm d\/}h}
{\int_{0}^{+\infty} C_{\rm n\/}^2(h)\, {\rm d\/}h}]^{3/5}
\end{equation}
Sarazin \& Tokovinin [\ref{ST}] give an empirical relation 
between $\overline{V}_{5/3}$
and the wind speeds measured at ground level and at 200\,mB pressure.
That relation has been verified at Paranal and Cerro Pachon in Chile, 
and later confirmed at San Pedro de Martir, Mexico:
\begin{eqnarray}\label{Eq:Vapprox}
\overline{V}_{5/3} \approx \max(V_{\rm g \/}, 0.4\,V_{\rm 200 mB\/})
\end{eqnarray}
At Paranal, $V_{\rm g \/}$ is measured by wind sensors at 30\,m height and 
$V_{\rm 200 mB\/}$ is estimated every 6 hours by the 
ECMWF [\ref{ECMWF}] (European Center for Medium Range Weather Forecast) 
through a global meteorological model which runs twice a day at 00\,UT and 12\,UT. 
This involves the assimilation of worldwide-collected data from 
radio soundings, satellite observations etc.\\
It appears from Fig.\,\ref{Fig:ASM_w} that the wavefront velocities derived with the
piston scope coincide with $0.4\,V_{\rm 200 mB\/}$, rather than 
$\overline{V}_{5/3}\approx \max(V_{\rm g \/}, 0.4\,V_{\rm 200 mB\/})$.
When the turbulence is contained in several layers, the measured structure function
is an average of single-layer structure functions as represented on Fig.\,\ref{Fig:SF}.
If these layers have different wind velocities and orientations,
the dispersion of the data points around the best-fitting structure function is large
and the resulting wavefront velocity is poorly constrained.
Accordingly, and in line with Fig.\,\ref{Fig:Profiles}, $V_{\rm ps \/}$ is derived with respectively
55\% and 10\% uncertainties during the first and second night of observations.

\item The coherence time (see Fig.\,\ref{Fig:ASM_t}) 
is a combination of the seeing and wavefront velocity,
thus it is essentially unconstrained during the first night. 
On the subsequent night, the values are consistent with those
derived through the two following methods: 
With MASS, $\tau_0$ is assessed from the scintillation through a 2\,cm diameter aperture.
MASS is not sensitive to the lower layers of turbulence ($<500$\,m), 
and, correspondingly, measures higher coherence times.
A second value of $\tau_0$ is obtained by combining DIMM seeing-values with 
measurements of the wind speed: 
$\tau_0  = 0.314 \; r_0  / \overline{V}_{5/3}$, where 
$\overline{V}_{5/3}$ is estimated by Eq.\,\ref{Eq:Vapprox}.
Since these values are obtained from distinct locations with different telescopes  
pointing at different stars, we do not expect them to coincide.
The results seem to suggest that the piston scope sees more turbulence than MASS and DIMM: 
While this is probable -- the piston scope is installed inside a dome at ground level,
whereas MASS and DIMM are placed on an open platform at 6\,m above the ground --
no definite conclusion is possible given the amount of data.
 
\end{itemize}

\section{Measurements at Dome\,C}\label{DomeC}

\begin{figure}[htbp]
\centering
\includegraphics[width=8.cm]{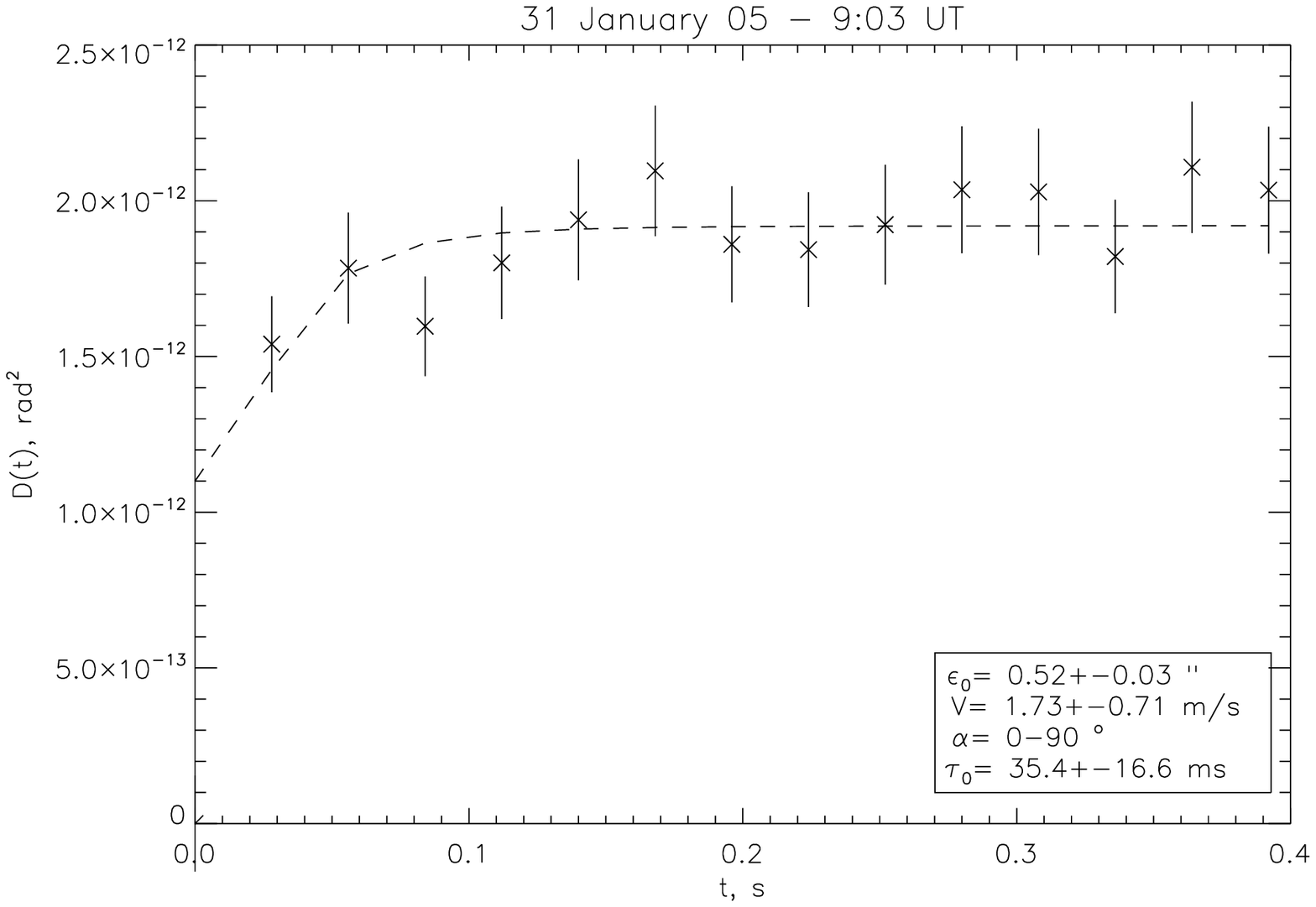}
\includegraphics[width=8.cm]{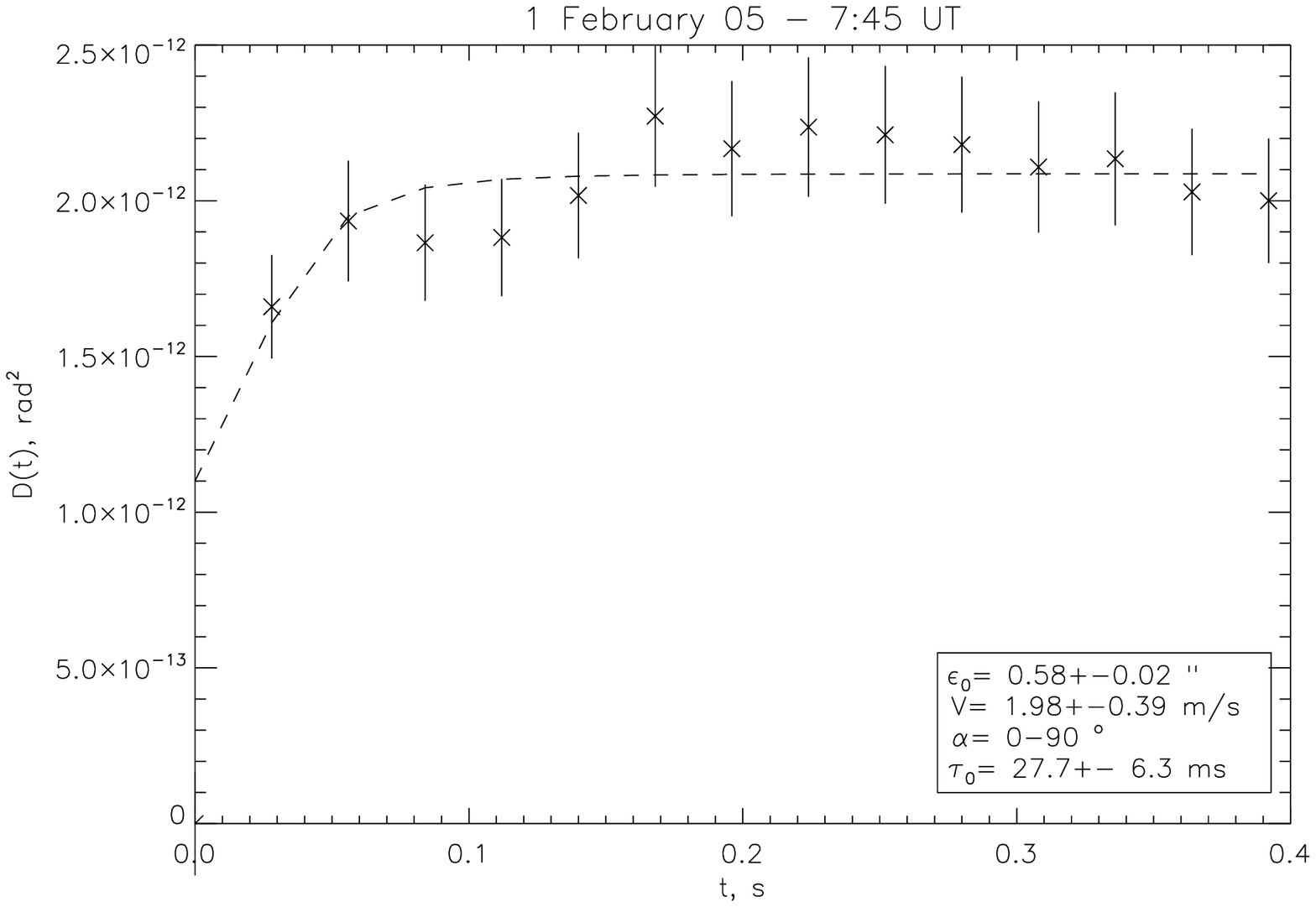}
\includegraphics[width=8.cm]{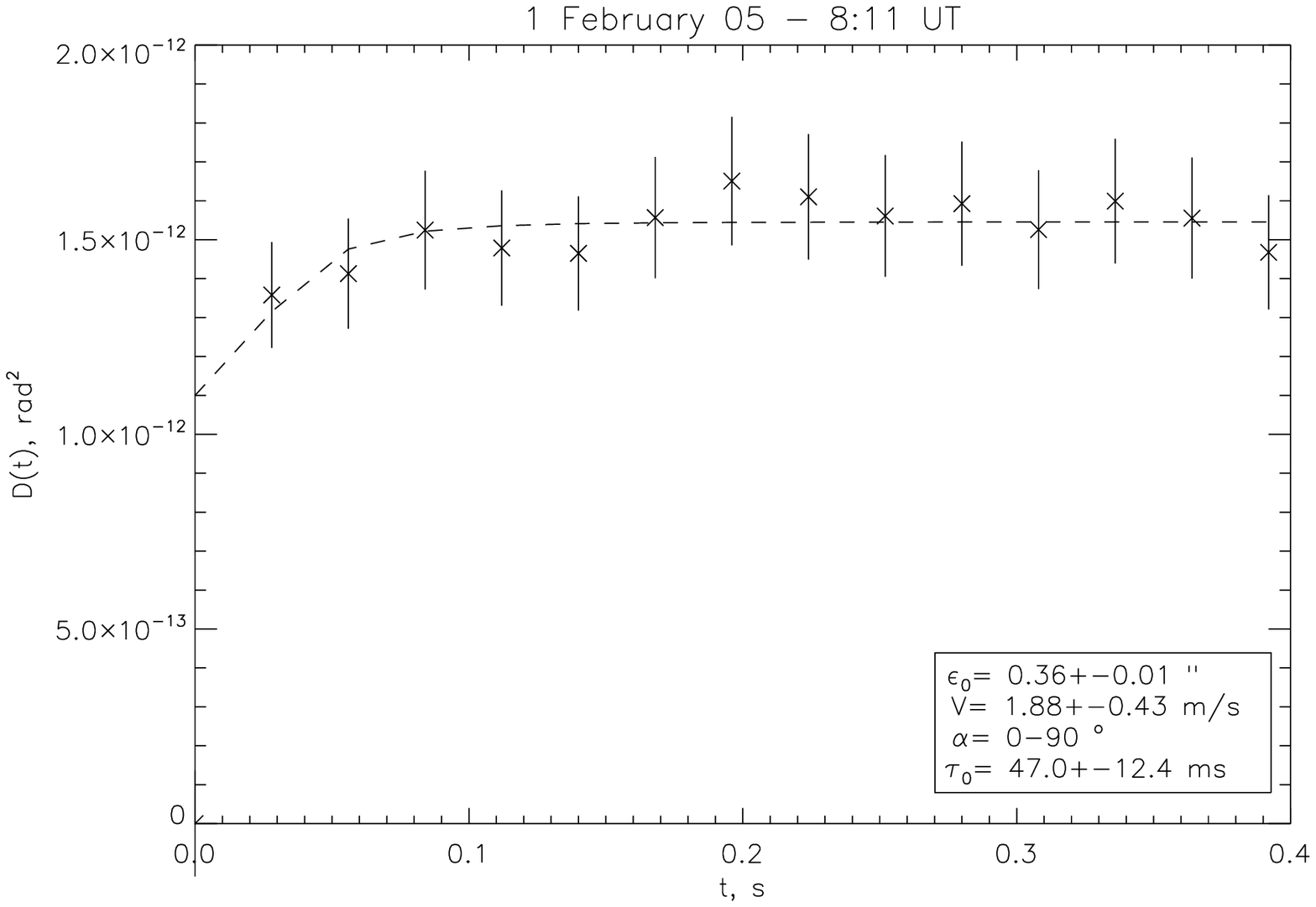}
\includegraphics[width=8.cm]{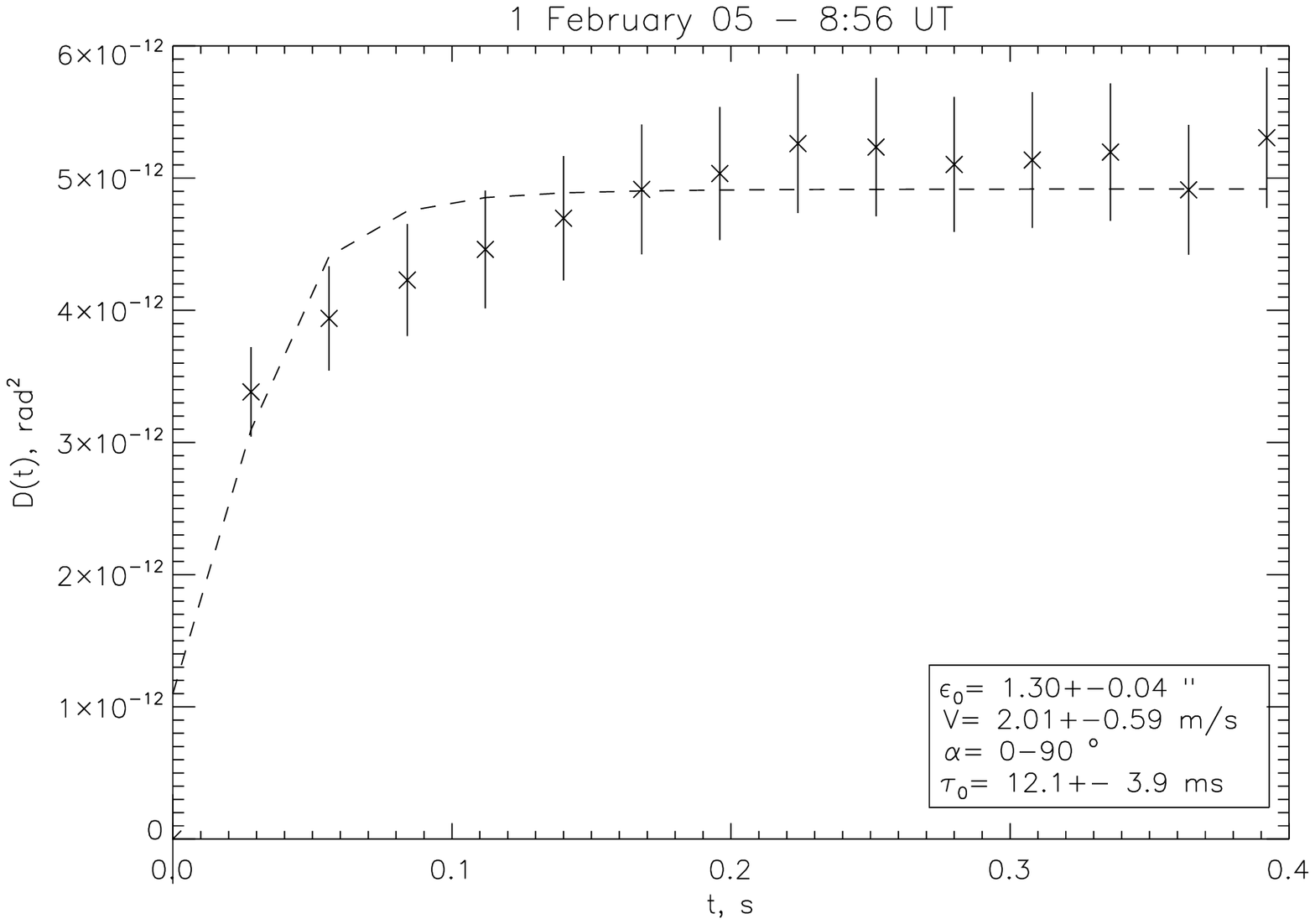}
\includegraphics[width=8.cm]{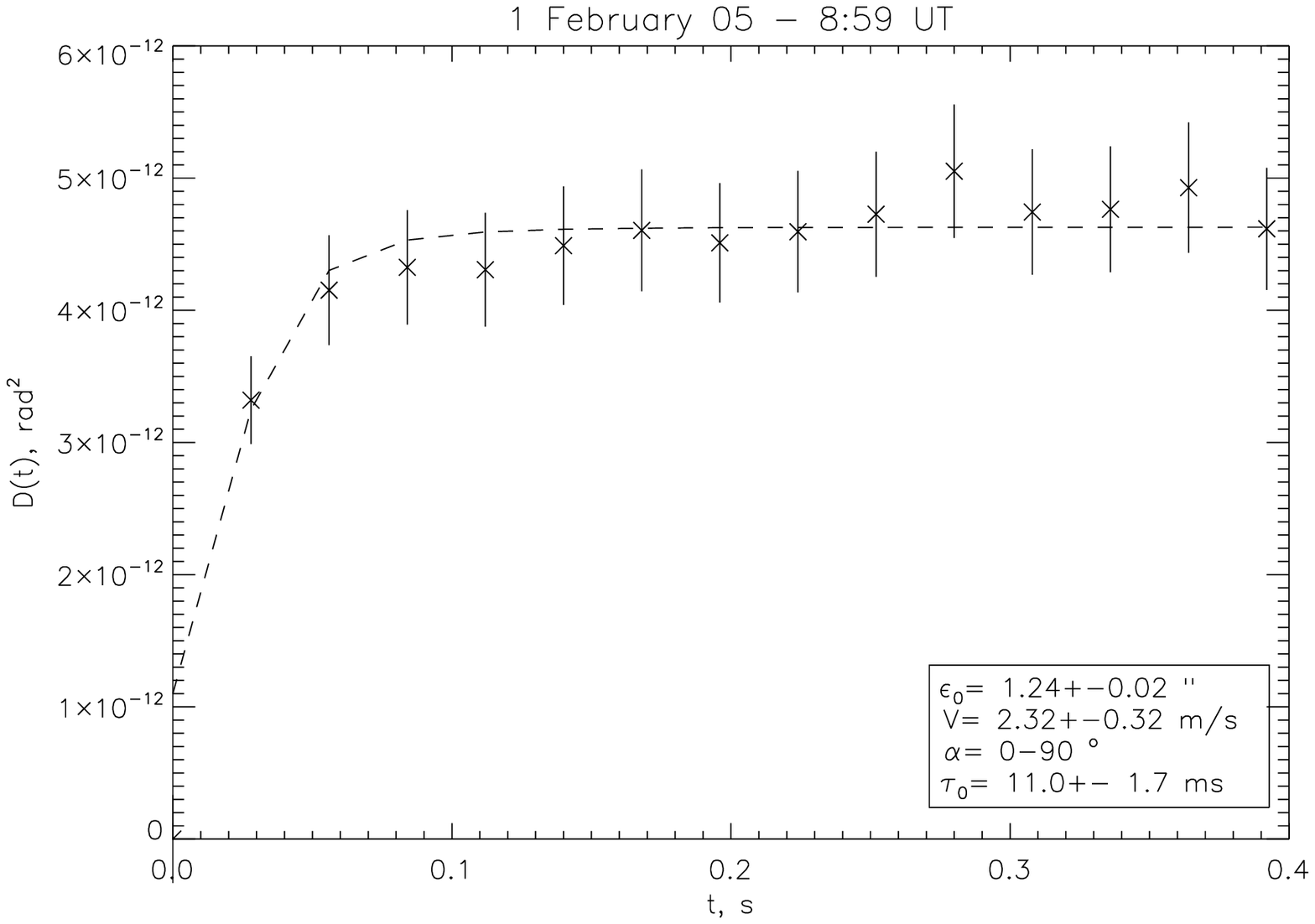}
\includegraphics[width=8.cm]{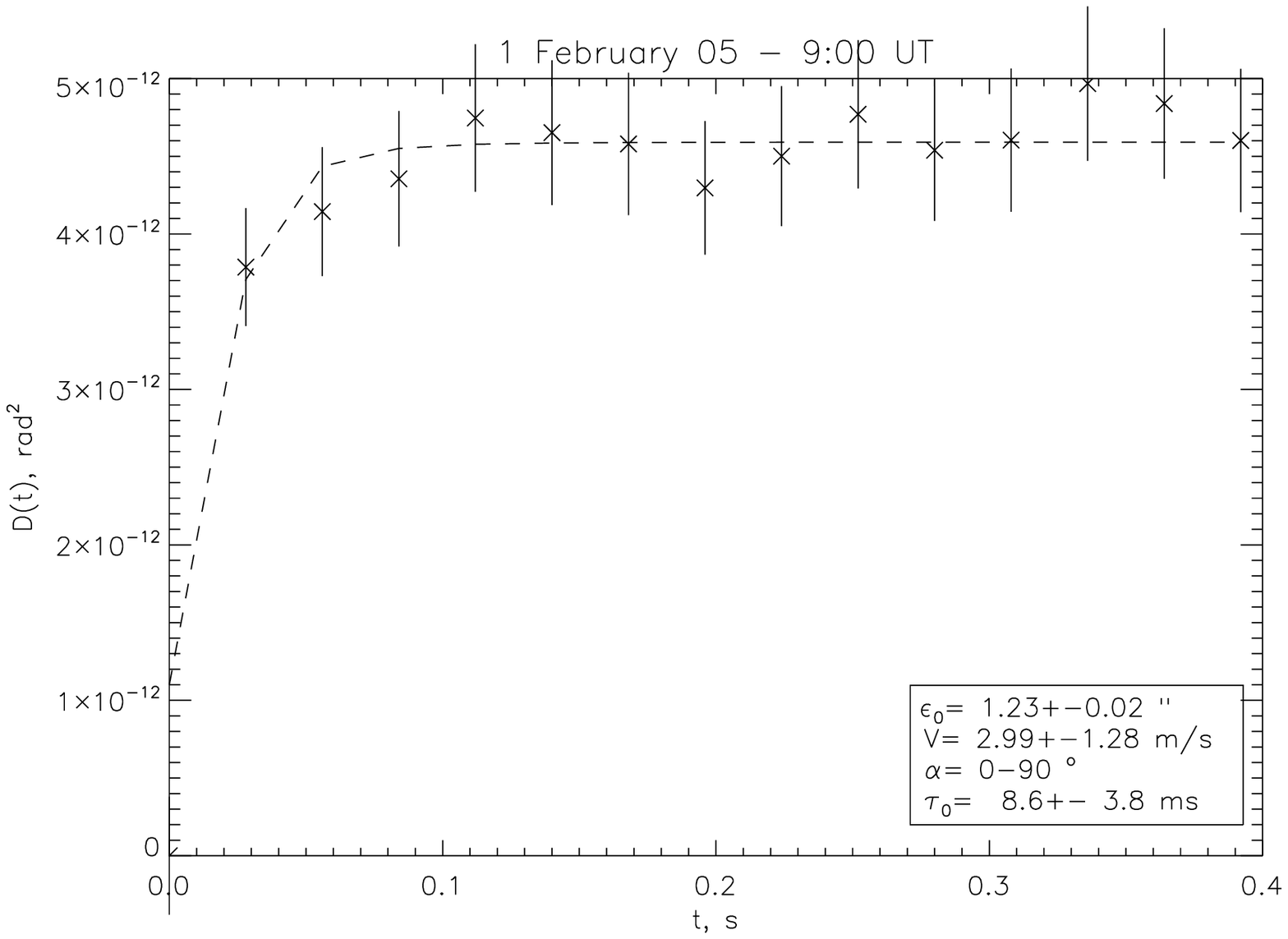}
\caption{ Atmospheric parameter values derived from measurements at Dome\,C.
}
\label{Fig:D}
\end{figure}

\begin{figure}[htbp]
\centerline{\includegraphics[width=18cm]{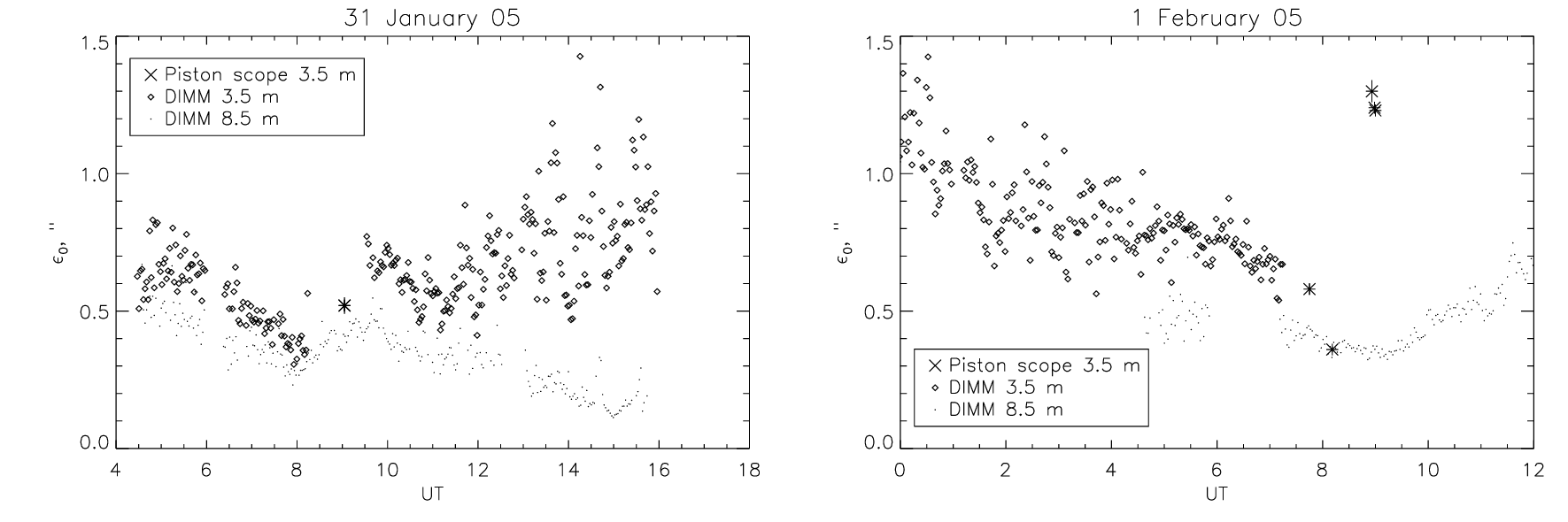}}
\caption{ Seeing values measured at Dome\,C with the DIMM and the piston scope.
}
\label{Fig:DIMM_s}
\end{figure}

Dome\,C is one of the summits on the Antarctic plateau with altitude 3233\,m. 
The station, which is jointly operated by France and Italy, is located 1100\,km inland from the French research station Dumont Durville and 1200\,km inland from the Italian Zuchelli station. 
Dome\,C is known as a site with extremely low wind speeds at high altitudes. 
Because of its elevated location and its relative distance from the edges of the Antarctic Plateau, Dome\,C does not experience the katabatic winds characteristic of the coastal regions of Antarctica. 
Hence the coherence times could be particularly high. 
Lawrence et al. [\ref{Lawrence}] have, during the Antarctic night, determined high-altitude turbulence parameters that are 2 to 3 times better than at mid-latitude sites.
Accordingly, they concluded that an interferometer located on Dome\,C might allow projects that would otherwise require instruments in space. 
The value of $\tau_0=7.9\,$ms obtained by Lawrence et al. was derived from measurements with the MASS instrument and hence, 
it does not take into account the turbulence below 500\,m
(see the MASS website for corresponding calibration studies [\ref{MASS}]).
Measurements of $\tau_0$  integrated over the whole atmosphere still need to be obtained.\\
In this context, similar measurements to those presented in Section\,\ref{Paranal}, have been performed at Dome\,C, Antarctica, on January $31^{\rm st\/}$ and February $1^{\rm st\/}$\,2005 at daytime.
For these measurements, {\it Canopus\/} was observed with a telescope of focal length 2.80\,m and a primary mirror of 0.28\,m, placed 3.5\,m above the ground. The entrance pupil was covered by a mask with two 0.06\,m diameter circular openings and centers 0.20\,m apart.
The observational set-up, as well as a first qualitative data analysis has been presented in a previous article [\ref{Kellerer1}].
The observations -- done with the available equipment --
were both spatially and temporally under sampled. 
During six sequences out of nine, it was nevertheless possible to place a lower limit 
equal to 10\,ms to the mean duration of correlation $t_{\rm c\/}$ of the fringe patterns. 
This was done by fitting an exponential curve onto the measured structure functions:
\begin{equation}
D_\phi(t)=D_\phi(t\gg t_{\rm c\/})\times(1-{\rm exp\/}(-(t/t_{\rm c\/})^{-5/3}))
\label{correlation}
\end{equation}
In Section\,\ref{Formalism}, the structure function has been related to the Fried parameter $r_0$ and to the velocity vector ${\bf V\/}$ in the case of a single turbulent layer,
using the Kolmogorov model of atmospheric turbulence. 
This relation has been tested on well sampled piston scope measurements recorded at Paranal (see Section\,\ref{Paranal}), and is now applied to re-analyze the data from Dome\,C.\\
We consider six out of nine sequences that were presented in the previous article. 
Images were taken every 28\,ms, with exposure times of 1, 2 or 3\,ms. 
Each sequence contains between 209 and 723\,images and, thus, lasts roughly 5 to 20\,s. 
Two sequences -- recorded on February $1^{\rm st\/}$ at 7:49\,UT and 9:41\,UT -- are not re-analyzed
because the central positions of the fringe pattern and of the combined Airy discs
are determined with too large uncertainties. 
In the previous article they were part of the three sequences during which the 
correlation time $t_{\rm c\/}$ was found to be less than 10\,ms. 
For the third such sequence, recorded on January $31^{\rm st\/}$ at 9:07\,UT,  the fringe pattern can be fitted but since 
the structure function reaches its asymptotic value at the first measurement point, it can not be compared to a theoretical curve.\\
As seen on Fig.\,\ref{Fig:D}, the structure functions reach their asymptotic value after the $4^{\rm th\/}$ to $5^{\rm th\/}$ data point.
The fit involves three free parameters $\epsilon_0, V, \alpha$ 
besides the white noise, 
$K$, that is approximately constant if the instrumental settings do not vary. 
The data obtained at Paranal from April $23^{\rm rd\/}$ to $24^{\rm th\/}$ yield:
$K=(3.2\pm0.7)\,10^{-14}$\,rad$^2$. 
To constrain the fit, $K$ is therefore fixed to the value that 
optimizes the global result of the six fitting procedures: $K=1.1\times 10^{-12}$\,rad$^2$.
\\
The derived parameter-values and uncertainties are indicated on Fig.\,\ref{Fig:D}. 
As specified in Section\,\ref{Paranal}, the uncertainties correspond to a two-fold increase in the
squared deviation of the data points to the theoretic structure function.
The values of the seeing are consistent with measurements by DIMM (Fig.\,\ref{Fig:DIMM_s}):
The difference in the estimates by the piston scope and the DIMM at 8.5\,m height,
resembles the scatter between the values estimated by the DIMM instruments at 3.5\,m and 8.5\,m,
and is due to ground layer turbulence.
In line with our previous qualitative analysis, coherence times are found to lie above 10\,ms
during the periods when five of the nine sequences were recorded.
\\
Note that the wind orientations are not constrained by the analysis.
To derive -- without continuous assessment of wind-direction profiles -- more accurate values of $\tau_0$, 
a parameter needs to be measured that is independent of the wind orientation. 
We have pointed out what appears to be a suitable new method in a previous article [\ref{Kellerer2}].

\section{Conclusions}
The atmospheric coherence time, $\tau_0$, is the crucial parameter for interferometers because it determines their sensitivity.
Yet, a simple method is still lacking to monitor the coherence time at different sites, 
and to decide where the future large interferometers ought to be built.
Does the piston scope fulfill this need?
To answer that question, we have related the measured quantity to parameters of the 
Kolmogorov model of turbulence. \\
It was found that due to its sensitivity to the wind direction 
the piston scope can be used to assess 
the wavefront vrelocity and the coherence time if, and only if,  
the whole turbulence is displaced along a single direction.
Since the single layer model is not a permanent feature on most sites,
the estimation of the coherence time is insecure.
This conclusion is supported by seven sequences of 1000 images, recorded with the 
piston scope at the observatory of Paranal in April 2005.
To determine the coherence time for any kind of atmospheric turbulence, 
a rotationally symmetric set-up has been proposed [\ref{Kellerer2}] 
and first measurements are planned.\\
The measurements performed at Dome\,C have been analyzed using the method here presented. 
Within the uncertainties due to low samplings, seeing angles are derived that 
coincide with simultaneous DIMM measurements. Mean wavefront speeds are found to be remarkably low. 
In agreement with a first qualitative analysis [\ref{Kellerer1}], the corresponding coherence times are determined to be superior to 10\,ms during five out of nine sequences.

%

\begin{figure}
\centering
\includegraphics[width=8cm]{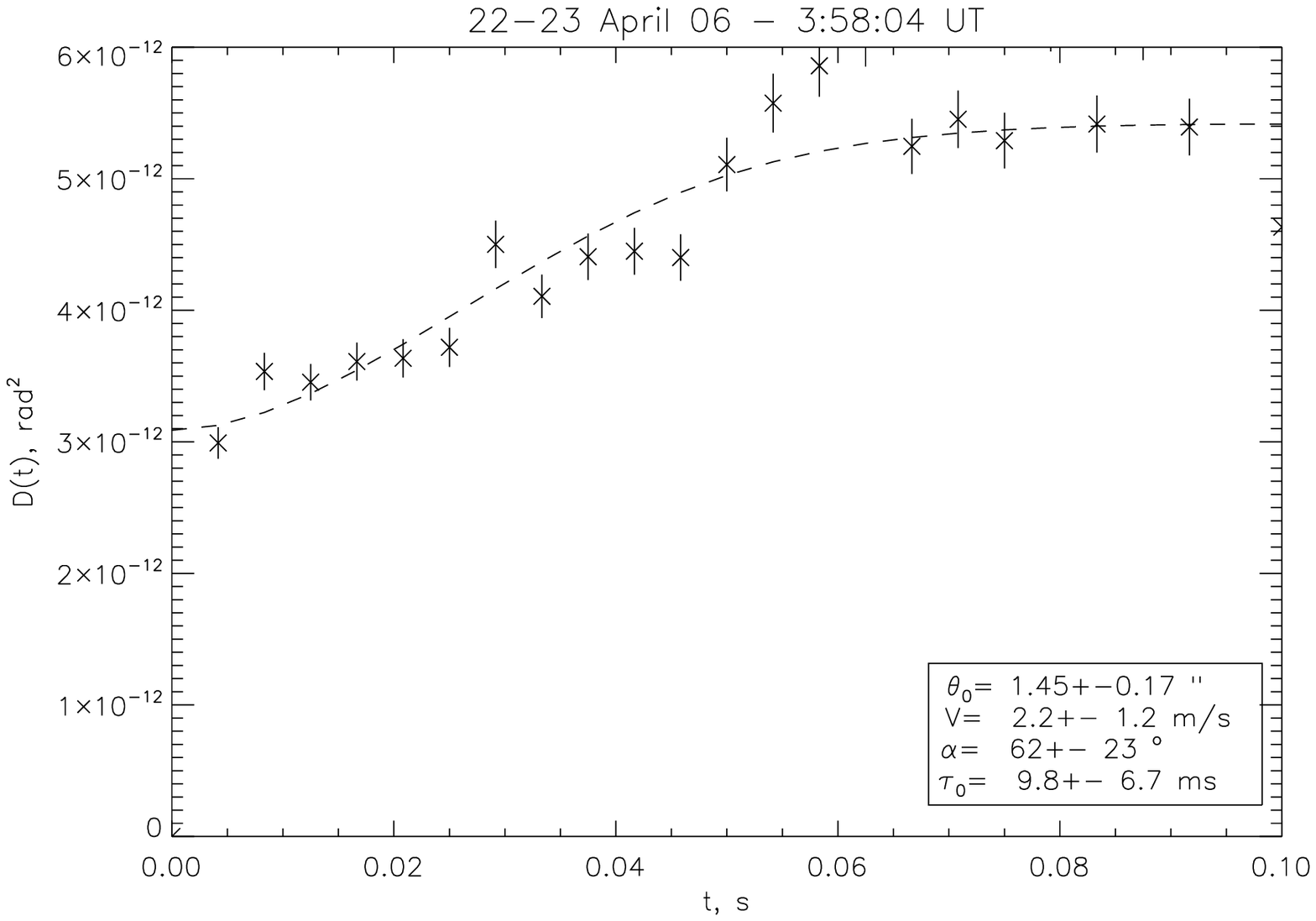}
\includegraphics[width=8cm]{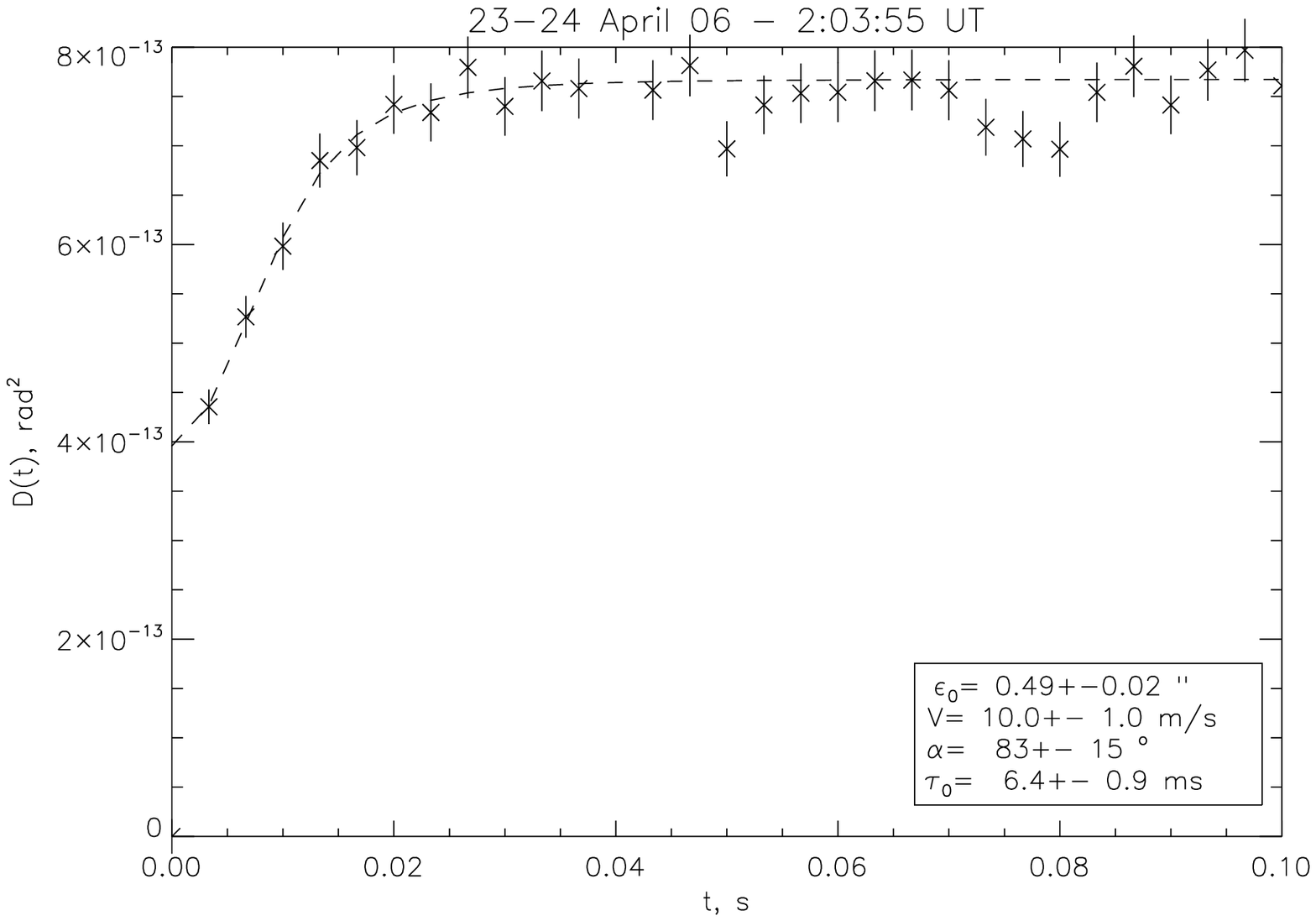}
\includegraphics[width=8cm]{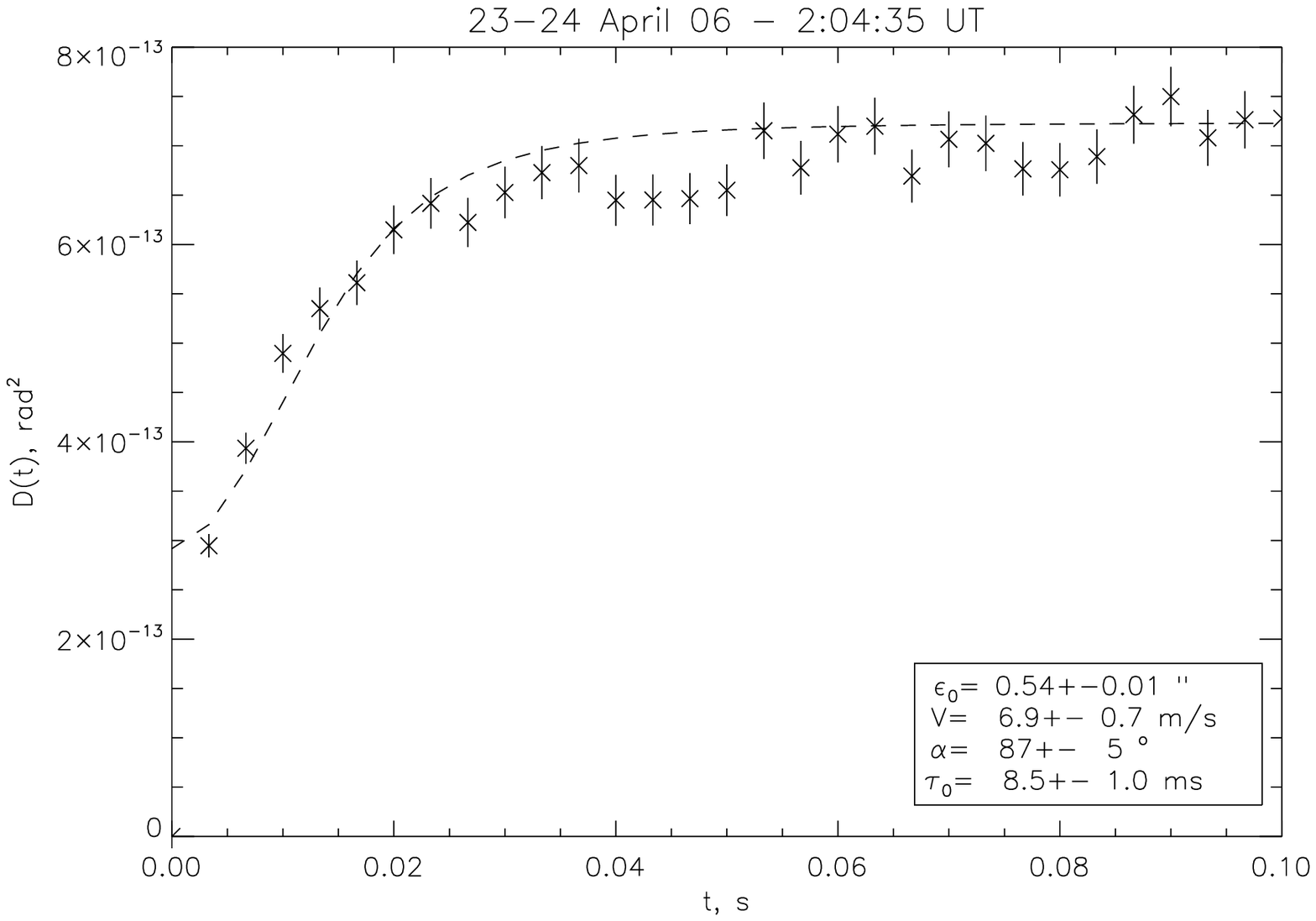}
\includegraphics[width=8cm]{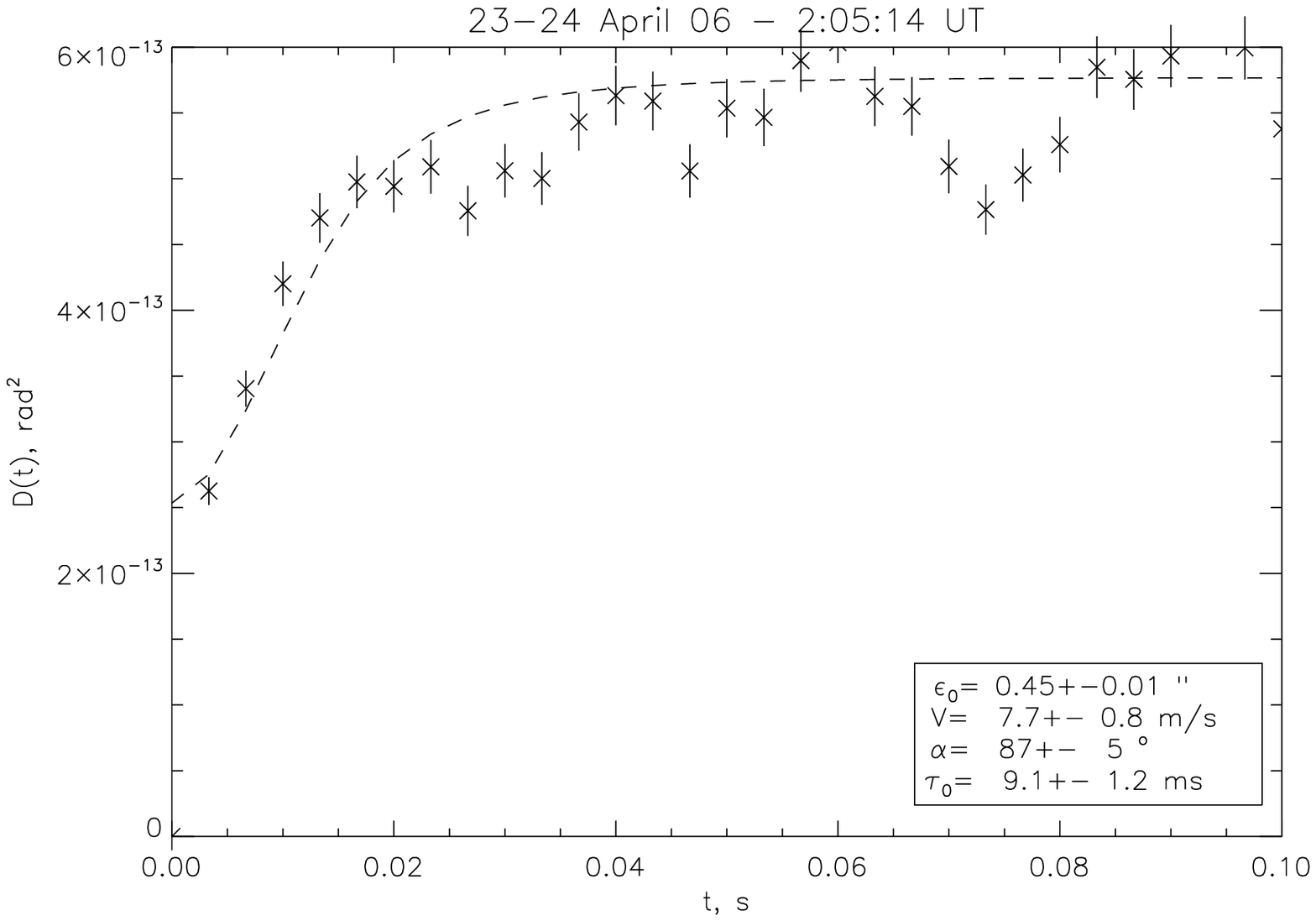}
\includegraphics[width=8cm]{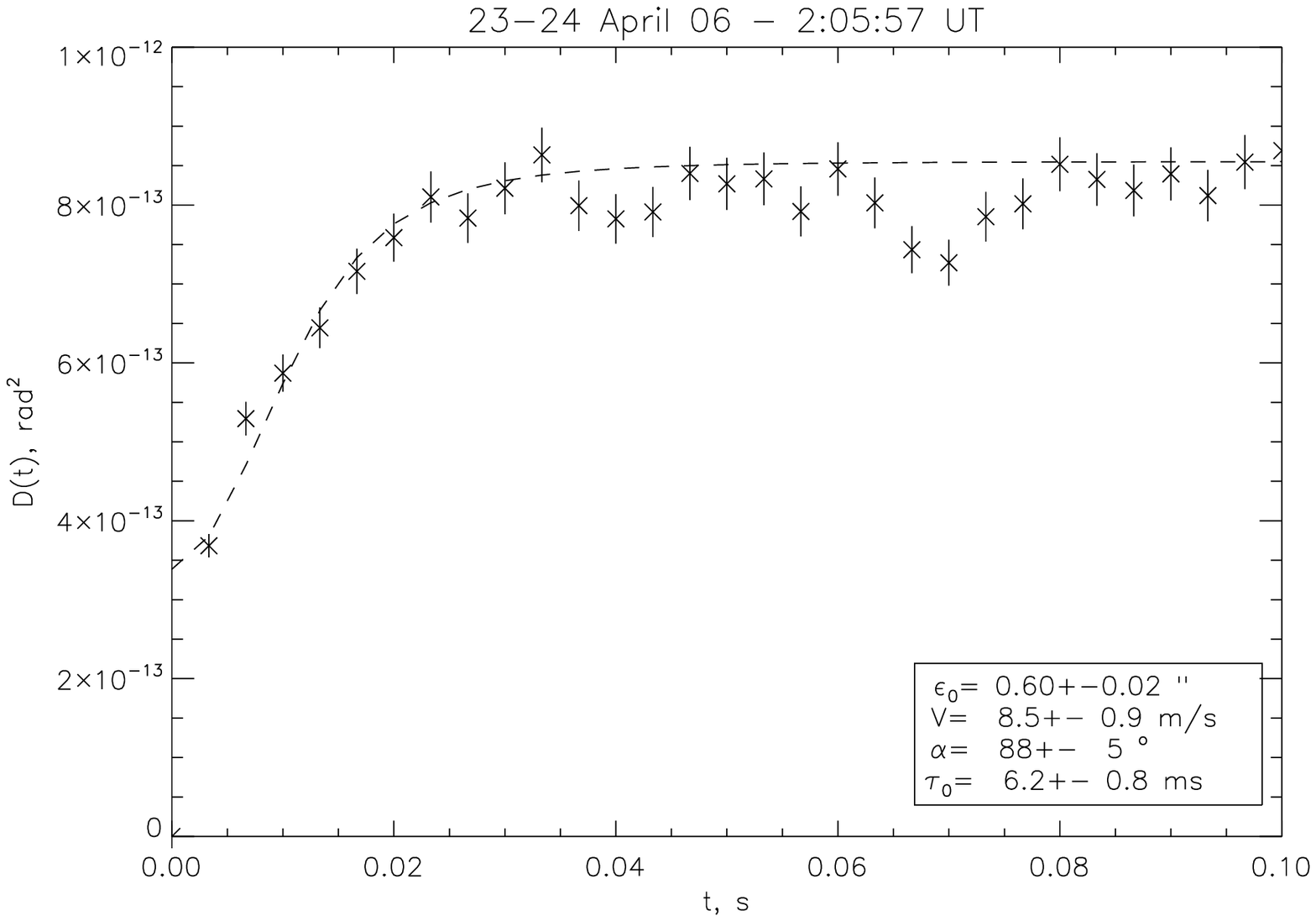}
\includegraphics[width=8cm]{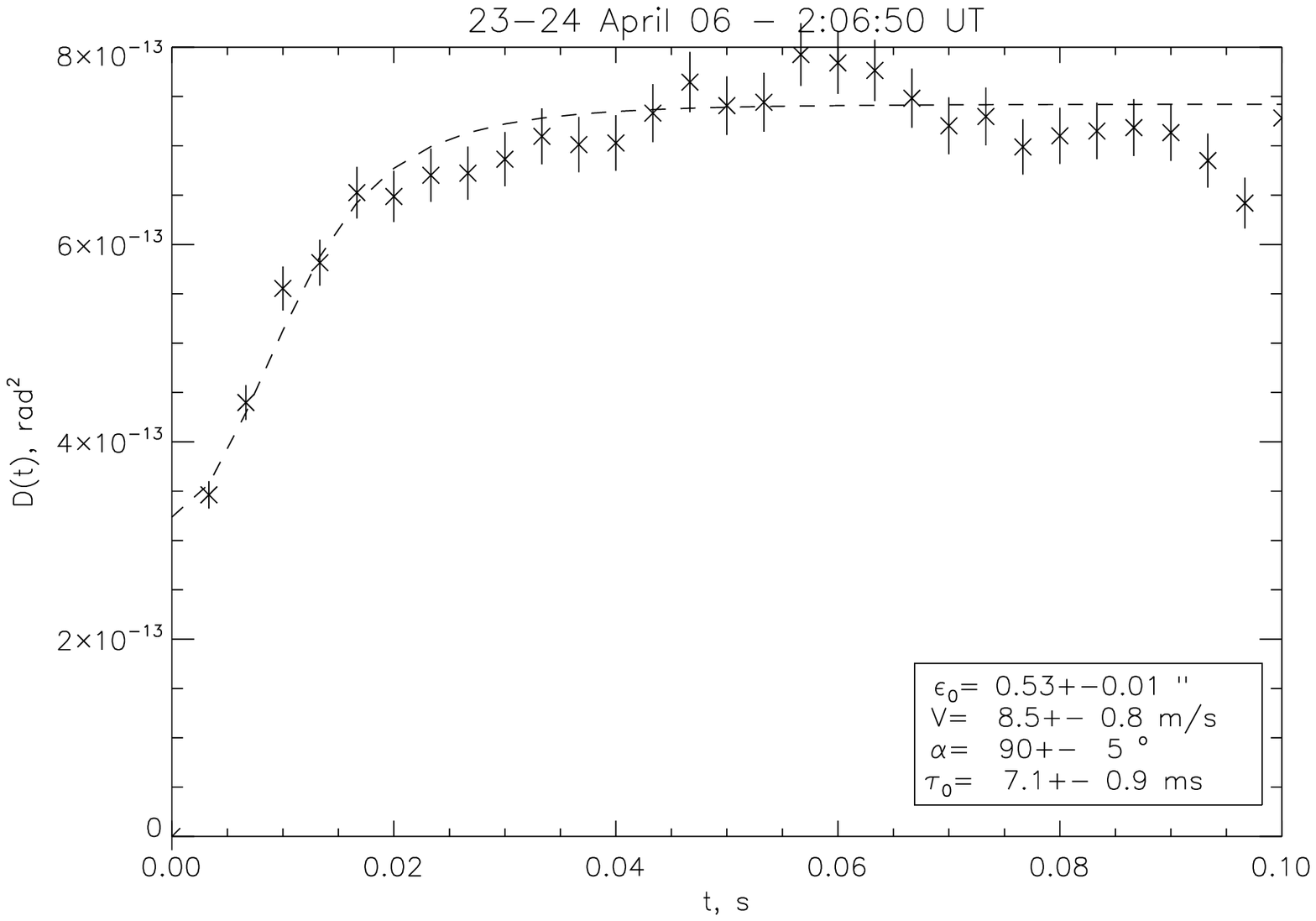}
\end{figure}

\begin{thebibliography}{99}

\bibitem{Kellerer1}\label{Kellerer1}
Kellerer, A., Sarazin, M., Coud\'e Du Foresto, V., Agabi, K., Aristidi, E., Sadibekova, T., ``A method of estimating time scales of atmospheric piston and its application at Dome C (Antarctica)", 
Applied Optics, {\bf 45}, 5709-5715 (2006)

\bibitem{Kellerer2}\label{Kellerer2}
Kellerer, A., Tokovinin, A., ``Atmospheric coherence time in interferometry: definition and measurement", A\&A, {\bf 461}, 775-781 (2007)

\bibitem{Conan}\label{Conan}
Conan, J.M., Rousset, G., Madec, P.Y., ``Wave-front spectra in high-resolution imaging through turbulence", J. Opt. Soc. AM. A, {\bf 12\/}, 1559-1570 (1995)

\bibitem{SLODAR}\label{SLODAR}
Butterley, T., Wilson, R., Sarazin, M., ``"Determination of the profile of atmospheric optical turbulence strength from SLODAR data", MNRAS, {\bf 369\/}, 835-845 (2006)

\bibitem{DIMM}\label{DIMM}
Sarazin, M., Roddier, F., ``The ESO differential image motion monitor" , A\&A, {\bf 227\/}, 294-300 (1990)

\bibitem{ST} \label{ST} 
Sarazin, M., Tokovinin, A.,
``The statistics of Isoplanatic Angle and Adaptive Optics Time Constant derived from DIMM data",
Beyond conventional adaptive optics, ESO Conference and Workshop Proceedings,
{\bf 58\/}, 321-328 (2001)

\bibitem{ECMWF} \label{ECMWF} 
European Centre for Medium-Range Weather Forecasts,
http://www.ecmwf.int/

\bibitem{Lawrence}\label{Lawrence}
Lawrence, J., Ashley, M., Tokovinin, A., Travouillon, T., 
``Exceptional astronomical seeing conditions above Dome C in Antarctica",
Nature, {\bf 431\/}, 278-281 (2004)

\bibitem{MASS}\label{MASS}
Tokovinin, A.,
National Optical Astronomy Observatory: MASS presentation,
http://www.ctio.noao.edu/$\sim$atokovin/profiler/ (2006)

\end{thebibliography}
\end{document}